%% file: main.tex
\documentclass[%
 reprint,
showpacs,
 amsmath,amssymb,
 aps,prl,
]{revtex4-1}
\usepackage{fancyvrb}
\usepackage[dvipsnames]{xcolor}
\usepackage{graphicx}
\usepackage{dcolumn}
\usepackage{bm}
\usepackage[mathlines]{lineno}

\DeclareMathOperator{\argmin}{argmin}

\begin{document}


\title{Random matrices and the New York City subway system}
\thanks{The authors thank Percy Deift for useful suggestions.  This work was supported in part by NSF-DMS-1303018 (TT) and NSF-OISE-1604232 (AJ).}%

\author{Aukosh Jagannath}
 \email{aukosh@math.utoronto.ca}
\author{Thomas Trogdon}%
 \email{ttrogdon@math.uci.edu}
\affiliation{%
 $^\dagger$Department of Mathematics, University of Toronto\\
 $^\ddagger$Department of Mathematics, University of California, Irvine
}%




\date{\today}

\begin{abstract}
We analyze subway arrival times in the New York City subway system.  We find regimes where the gaps between trains exhibit both (unitarily invariant) random matrix statistics and Poisson statistics.  The departure from random matrix statistics is captured by the value of the Coulomb potential along the subway route.  This departure becomes more pronounced as trains make more stops.
\end{abstract}

\maketitle



The bus system in Cuernavaca, Mexico in the late 1990s has become a canonical physical system that is well-modeled by random matrix theory (RMT) \cite{Krbalek2000,Krbalek2001,Krbalek2003,Krbalek2008}.  This bus system has a built-in, yet naturally arising, mechanism to prevent buses from arriving in rapid succession.  
If a driver arrives at a stop just after another bus on the same route, there will be few fares to collect so the self-employed drivers introduced a scheme, using a cadre of observers along each route, to space themselves apart so as to maximize the number of fares they collect.  Without this interaction, and mutual competition, one should expect that bus arrivals would be Poissonian \cite{OLoan1998}.  While the New York City subway (MTA) system has a different, globally controlled, mechanism to space trains to eliminate collisions, much of the MTA system remains under manual control \cite{Dougherty2002}. In this letter, we compare the predictions and results from Cuernavaca, Mexico with the MTA system. 

In particular, the authors in \cite{Krbalek2000} noted that if one stood at bus stop in Cuernavaca, Mexico, near the city center, and recorded the set $T$ of times between successive buses then for $\tau = T/ \langle T \rangle$
\begin{align}\label{e:gap}
\frac{\#\{ s \in \tau : s \leq t\}}{ \# \tau } \approx \int_0^t \rho(s)  ds, ~ \rho(s) = \frac{32}{\pi^2} s^2 e^{-\frac{4}{\pi} s^2},
\end{align}
where $\langle \cdot \rangle$ represents the sample mean and the function $\rho(s)$ is known as the ($\beta = 2$) \emph{Wigner surmise} (WS) \cite{MehtaRM}. This is the approximation of Eugene Wigner for the asymptotic ($N \to \infty$) gap distribution for successive eigenvalues in the bulk of an $N \times N$ GUE (Gaussian Unitary Ensemble) matrix \footnote{A GUE matrix is a Hermitian matrix with iid standard complex Gaussian entries, up to the symmetry condition.}.  This is computed by considering the $2 \times 2$ case.  This approximation of Wigner agrees surprisingly well with the true limiting distribution as $N \to \infty$ \footnote{A numerical calculation using Fredholm determinants reveals that the KS distance is less than $5 \times 10^{-3}$. }.

The authors in \cite{Krbalek2000} consider another statistic called the \emph{number variance}.  Fix a time $T_0$ and consider the time interval, $[T_0,T]$, for $T_0 \leq T \leq T_1$.  Let $n(T)$ be the number of buses (or subway trains) that arrive in this time interval.  Once one has made many statistically independent observations of $n(T)$, the number variance is computed by
\begin{align}\label{e:num_var_comp}
N(t) = \langle (n(T) - \langle n(T) \rangle )^2 \rangle, \quad T = T_1 \langle n(T_1) \rangle^{-1}t.
\end{align}
This normalization is made so that $\langle n(T) \rangle \approx t$.  The aysmptotic prediction from RMT is 
\begin{align}\label{e:num_var} N(t) \approx \frac{1}{\pi^2}(\log 2 \pi t + \gamma + 1) \end{align}
where $\gamma$ is the Euler constant \cite{MehtaRM}.   This prediction is verified for the Cuernavaca bus system in \cite{Krbalek2000}.  A physically-motivated model for the bus system was presented in \cite{Baik2006} for which \eqref{e:gap} and \eqref{e:num_var} hold.

In this letter,  we observe that \eqref{e:gap} and \eqref{e:num_var} hold on a subset of the MTA system. We also find Poisson statistics within the MTA (which are also found in Puebla, Mexico \cite{Krbalek2003}).  For example, the southbound \#1 train in northern Manhattan exhibits RMT statistics but the northbound \#6 train exhibits Poisson statistics in the middle of its route. We also show that the train gap statistics tend to deviate more from RMT statistics as more stops are made. To quantitatively determine Poisson statistics versus RMT statistics we make the following ansatz for the (normalized to mean one) gap density function for $u \in [0,1]$
\begin{align*}
p(s;u) := \int_0^s \frac{\rho(x(1-u)^{-1})}{1-u} (1 - e^{(x-s)/u}) dx.
\end{align*}
This is the density for the convex combination of an independent exponential and a WS random variable. A similar ansatz was used in \cite{Abul-Magd2007a} for an analysis of car spacing statistics. Using the Kolmogorov--Smirnov (KS) statistic we choose $u$ to fit this distribution to the data.  A small value of $u$, combined with a small KS value indicates RMT statistics.  A value of $u$ near unity, and a small KS value indicates Poisson statistics.  We note that this transition (from RMT to Poisson) is also seen within RMT as the bandwidth of a Hermitian random matrix shrinks \cite{Shcherbina2014}.

\paragraph{Data collection.} Our data is obtained from the MTA Real-Time Data Feeds \cite{MTAData} that allow the user to obtain real-time train arrival times for many stations in the MTA system.  Thus, our analysis has an advantage over that in \cite{Krbalek2000} because the statistics of every station in the data feed can be analyzed.  The stations can then be classified into those exhibiting RMT statistics, Poisson statistics or neither.  Using the latitude and longitude coordinates of each station, which the MTA also provides, we can estimate the arc length of the subway track and analyze spatial distances. This is a component in our Coulombic analysis below.

We analyze the arrival times for the \#1 and \#6 trains.  These trains operate on separate lines.  The \#1 train runs both northbound and southbound between Manhattan and the Bronx.  The \#6 train runs both northbound and southbound, also between Manhattan and the Bronx.  The stations at which the \#1 train stops are labeled with integers between $101$ and $142$ \footnote{A table to convert from station number to station name can be found the supplemental materials.}, increasing from north to south.  The same is true of the stations for the \#6 train with integers ranging between $601$ to $640$. Our data set consists of \#1 and \#6 train arrival times in seconds at all stations obtained on 48 days (39 weekdays) during the summer and fall of 2016.  We only consider arrivals that occur between 8am and 6pm on weekdays.  For each station we have approximately 3500 arrivals. The MTA system keeps a minimum spacing between trains, unlike the Cuernavaca bus system.  To account for this, we subtract 90 seconds from every train gap.  This number could be treated as a fitting parameter, but we keep it fixed.  This leads to a small number of negative gaps. Then if $T$ is the collection of observed gaps (in seconds) define $\tau = (T-90)/\langle T - 90\rangle$ to be the normalized train gaps.

\begin{figure}[tb]
\includegraphics[width = .9\linewidth]{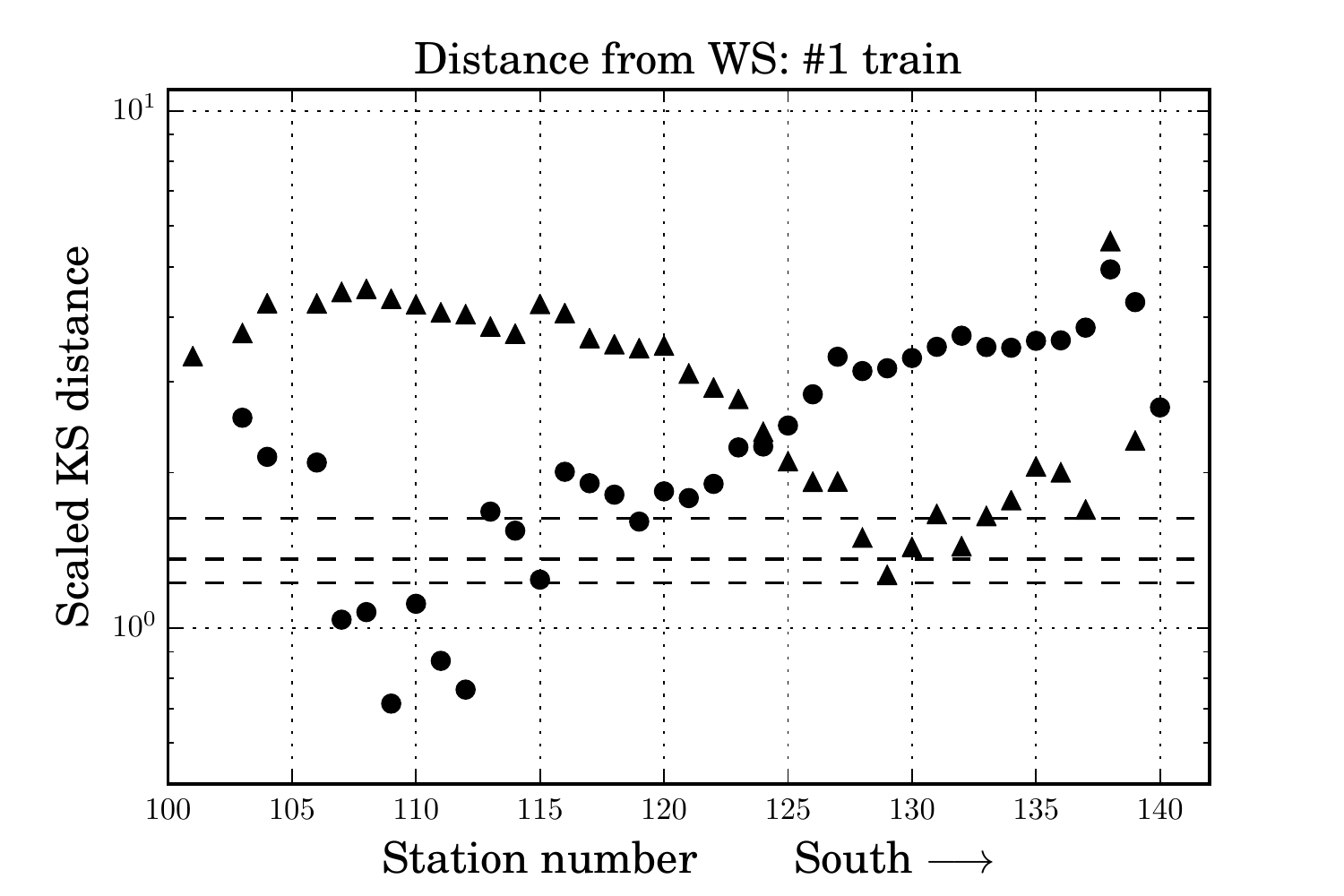}
\includegraphics[width = .9\linewidth]{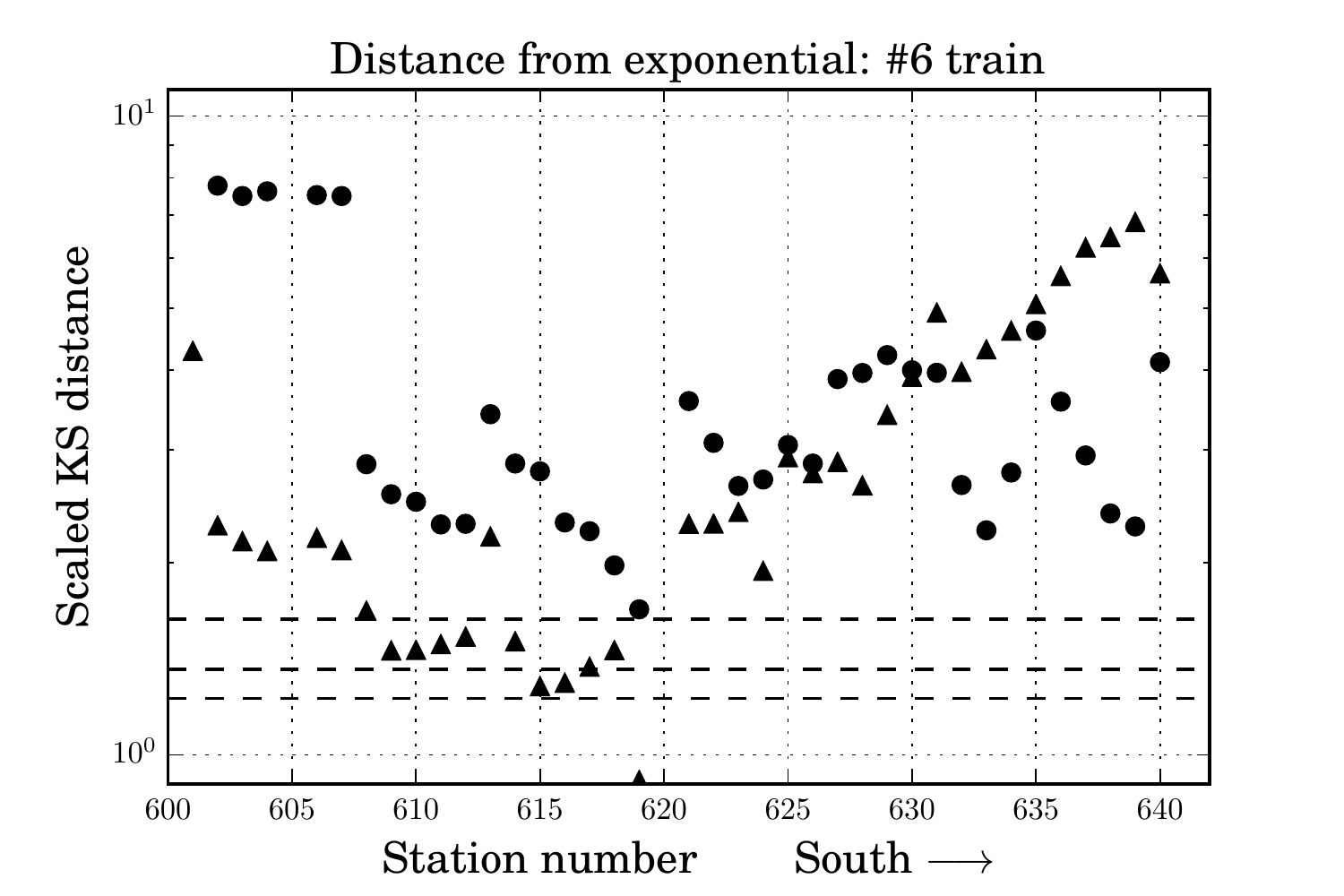}
\caption{\label{f:KS-test} The KS test for the \#1 train (top, $u = 0$) and the \#6 train (bottom, $u = 1$). Circles and triangles represent southbound and northbound trains, respectively. The dotted lines in order of decreasing height represent the significance levels  $\alpha = 0.01, 0.05, 0.1$.  Stations that lie below a line pass the associated KS test.  }
\end{figure}

\paragraph{The Kolmogorov--Smirnov test.}  Define the KS statistic \footnote{$\# \tau$ is the cardinality of the set $\tau$.}
\begin{align*}
\mathrm{KS}(u,\tau) &:= \sup_{t \in \mathbb R} \left| \frac{\#\{ s \in \tau : s \leq t\}}{ \# \tau }  - \int_0^t p(s;u) ds \right|.
\end{align*}
For $u = 0$, the null hypothesis is that the normalized gaps are distributed according the WS and for $u = 1$, the null hypothesis is that the gaps are exponentially distributed.  The KS test supposes that the samples are independent.  From our data we obtain successive gaps which contain repeated data from the same train and are clearly not independent.  To approximate independence, we only retain every fifth gap and we perform the KS test with approximately 700 samples.  We consider the significance levels $\alpha = 0.01, 0.05, 0.1$ (low, moderate and high significance, resp.). It follows from \cite{Kolmogorov1933,Smirnov1948} that the null hypothesis cannot be rejected if ($u = 0,1$)
\begin{align*}
\sqrt{\# \tau} \mathrm{KS}(u,\tau) < 1.63 ~~~\text{ when }~~ \alpha = 0.01,\\
\sqrt{\# \tau} \mathrm{KS}(u,\tau) < 1.36 ~~~\text{ when }~~ \alpha = 0.05,\\
\sqrt{\# \tau} \mathrm{KS}(u,\tau) < 1.225 ~~\text{ when }~~ \alpha = 0.10.
\end{align*}
In Fig.~\ref{f:KS-test}, we plot this scaled KS test statistic for every station on the northbound and southbound \#1 and \#6 trains.  In particular, we find with high statistical significance ($\alpha = 0.10$) that six stations (107, 108, 109, 110, 111, 112) for the southbound \#1 train pass the $u = 0$ KS test. If $\alpha$ is reduced, more stations pass the test.  Similary, for the northbound \#6 train, one station passes the $u = 1$ KS test with high significance (619) and a total of three (615, 616, 619) stations pass the same test with moderate significance.

\begin{figure}[tb]
\includegraphics[width = .9\linewidth]{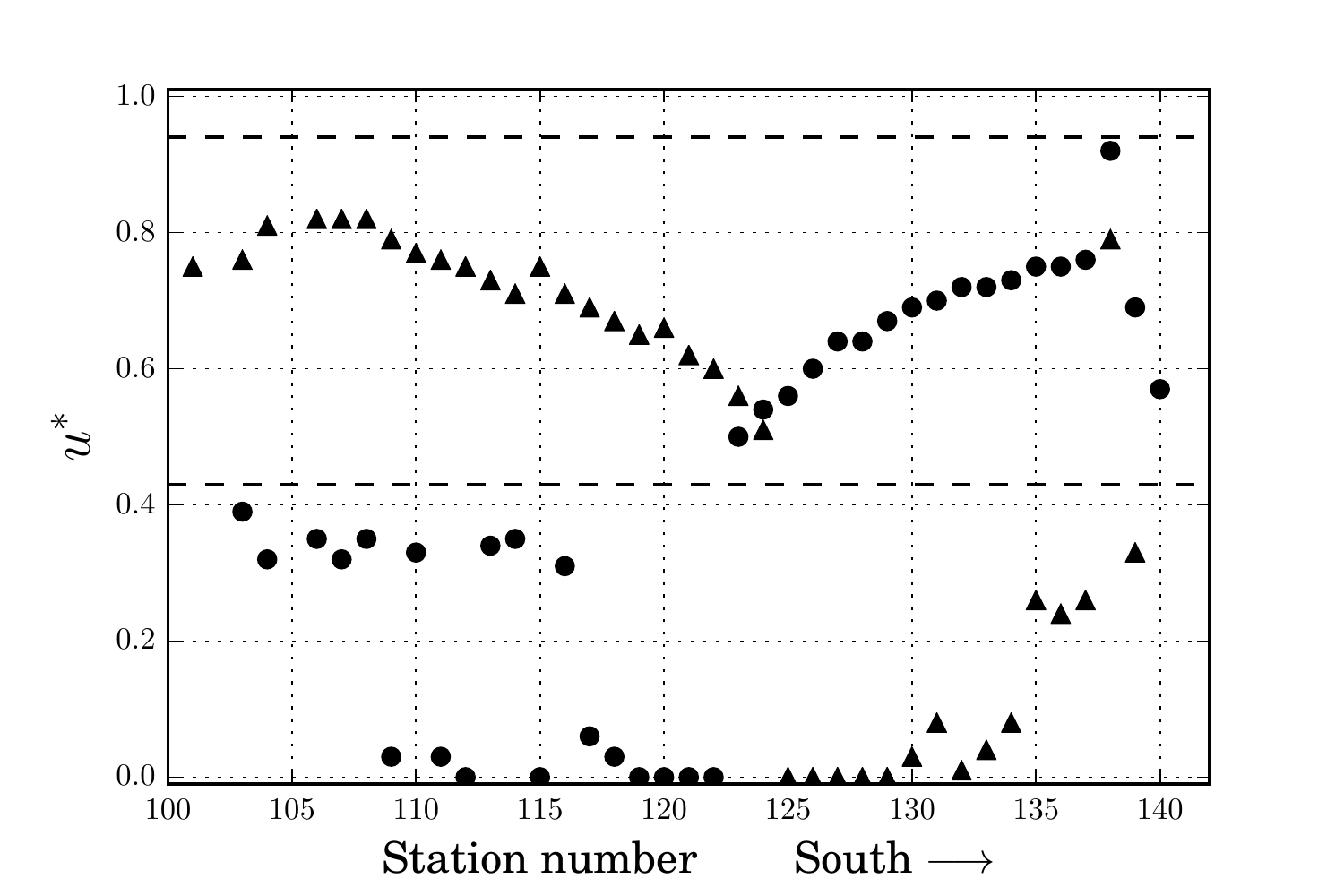}
\includegraphics[width = .9\linewidth]{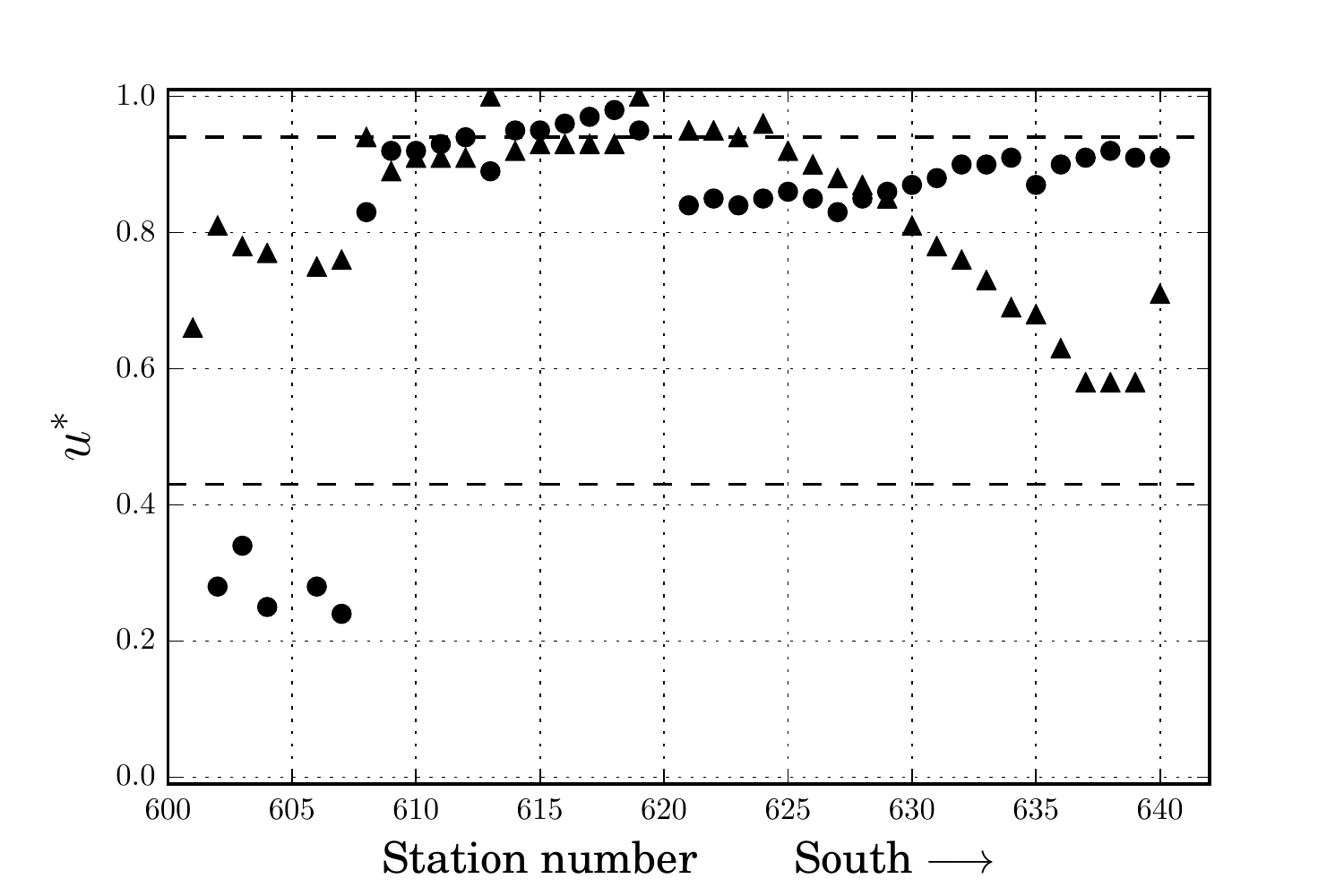}
\caption{\label{f:ustar} The KS fit for the \#1 train (top, $u = 0$) and the \#6 train (bottom, $u = 1$). Circles and triangles represent southbound and northbound trains, respectively. The dotted lines give the $u^* = 0.43$ and the $u^* = 0.94$ thresholds.  Values of $u^*$ above $0.94$ indicate Poisson statistics and values of $u^*$ below $0.43$ indicate RMT-like statistics.}
\end{figure}

\paragraph{A Kolmogorov--Smirnov fit.}   The value of $u^*$ of $u$ that fits the data best is given by
\begin{align*}
u^* &:= \underset{0 \leq u \leq 1}{\argmin} ~\mathrm{KS}(u,\tau).
\end{align*}
For every collection of normalized gaps $\tau$ this gives an optimal value $u^*$.  Recalling that our sample sizes are approximately $700$, we find that for $u < 0.43$ the KS test with moderate significance (comparing with $u = 0$) is passed.  For $u > 0.94$ we find that the KS test with moderate significance is passed when comparing with $u = 1$.  Stations with $u^* < 0.43$ are considered to exhibit RMT-like statistics and stations with $u^* > 0.94$ are considered to exhibit Poissonian statistics.  The values of $u^*$ for each station and train is given in Fig.~\ref{f:ustar}. These results should be compared with Fig.~\ref{f:KS-test} to ensure significance.  This presents further evidence that train gaps on the \#1 train are RMT-like and those on the \#6 train are Poissonian.

\begin{figure}[tb]
\includegraphics[width = .9\linewidth]{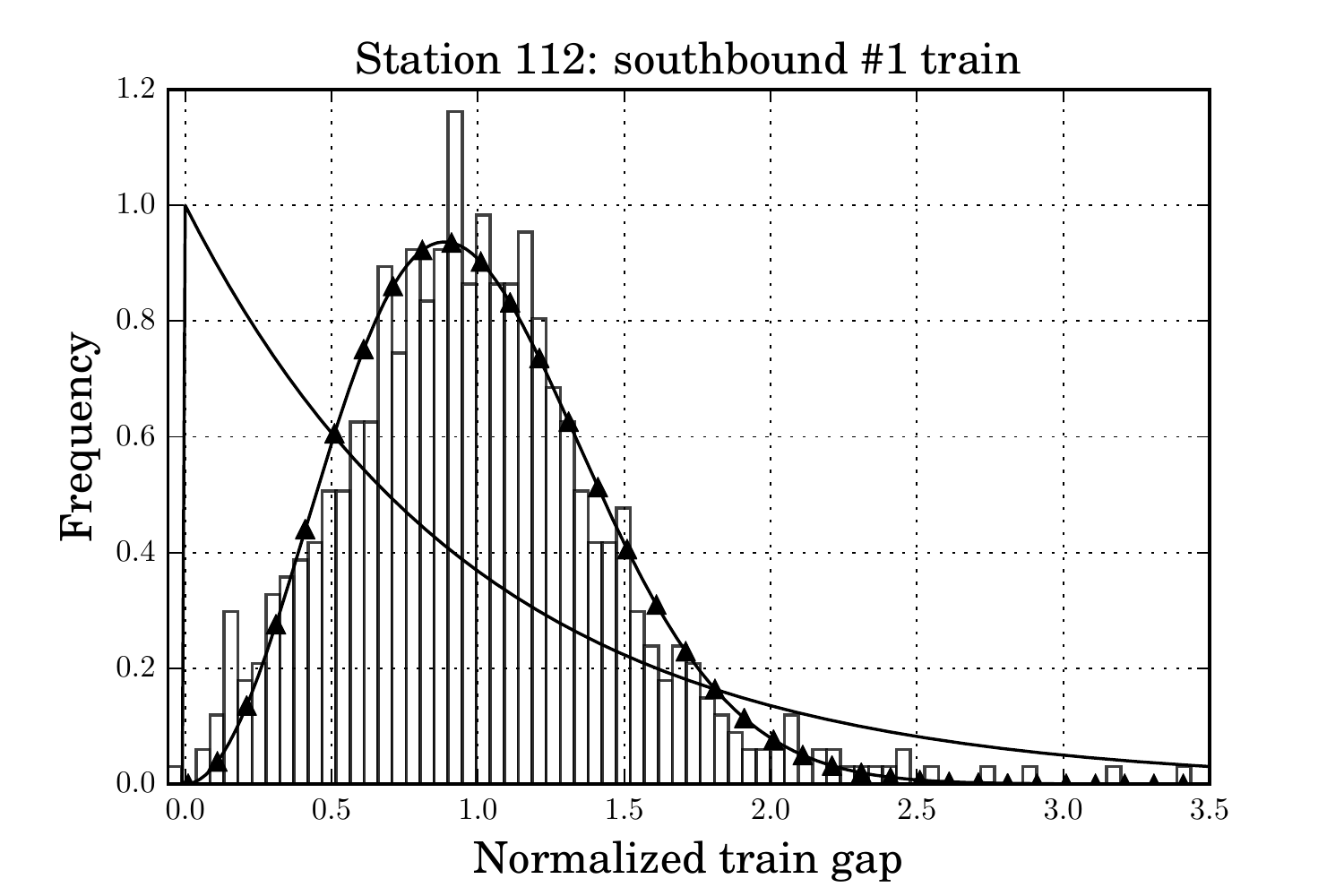}
\includegraphics[width = .9\linewidth]{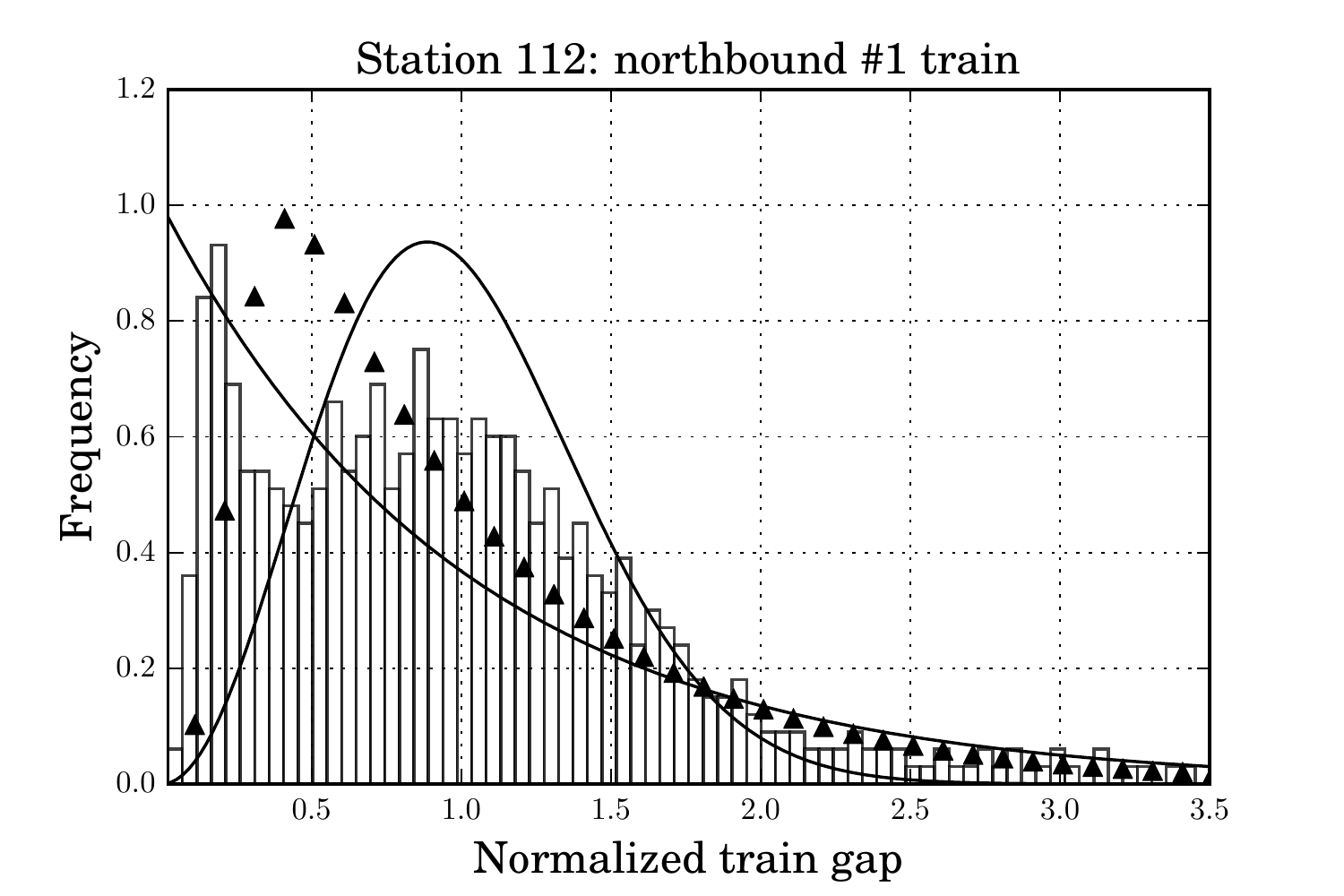}
\caption{\label{f:112} The normalized train gap histograms for the northbound (bottom) and southbound (top) \#1 trains at station 112. The solid curves give the exponential and WS density. The triangles represent the best-fit density $p(s;u^*)$.  The southbound train exhibits (highly significant) RMT statistics and our ansatz that determined $p(s;u^*)$ is not sufficient to capture the behavior of northbound trains.}
\end{figure}

\begin{figure}[t]
\includegraphics[width = .9\linewidth]{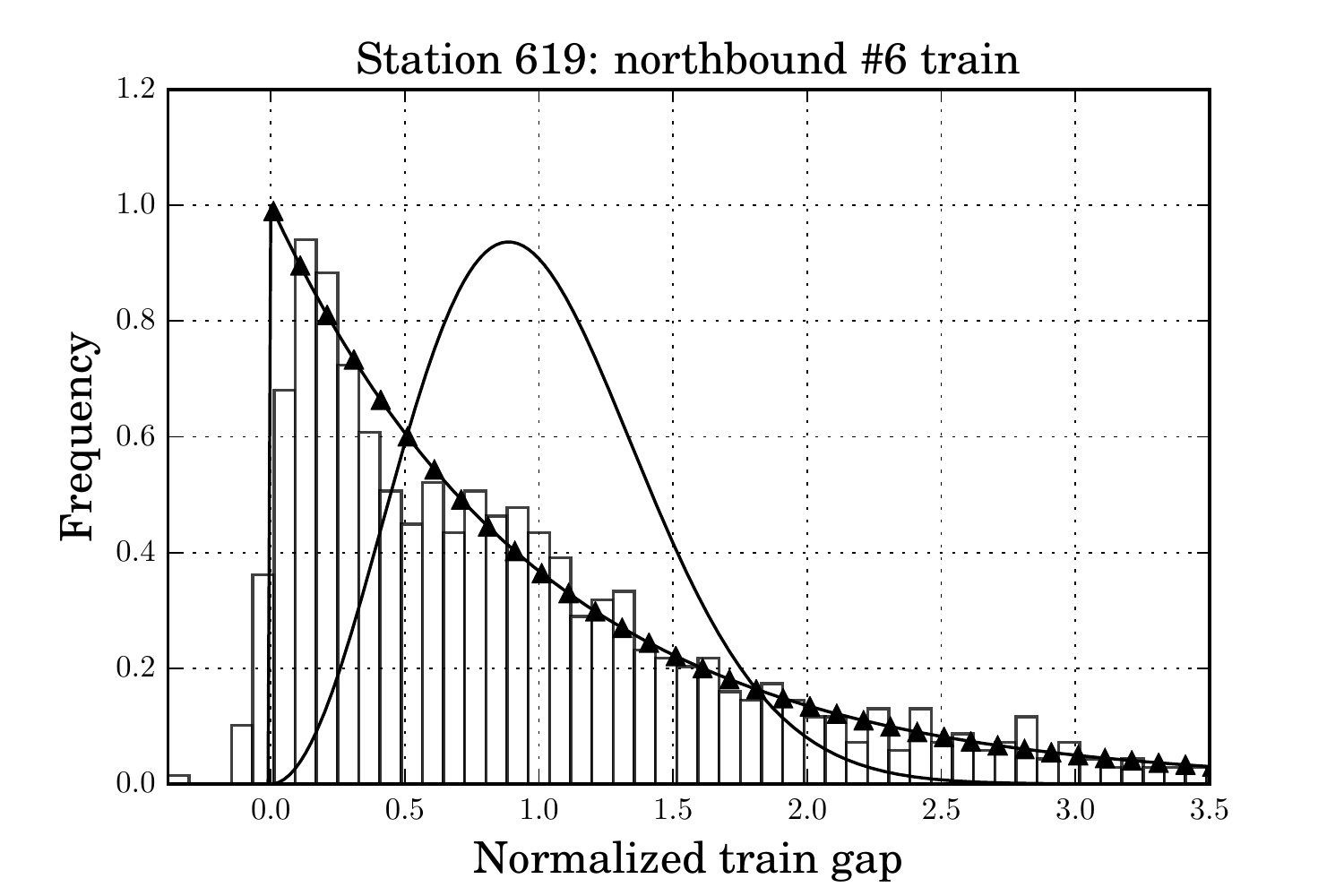}
\caption{\label{f:619} The normalized train gap histograms for the northbound \#6 trains at station 619. The solid curves give the exponential and WS density. The triangles represent the best-fit density $p(s;u^*)$.  This station exhibits (highly significant) Poissonian statistics.}
\end{figure}

We choose station 112 and station 619 to examine in more detail.  We display the normalized train gap histogram for both northbound and southbound trains at station 112 in Fig.~\ref{f:112}.  It is clear (and indeed highly statistically significant) that the southbound train gaps exhibit RMT statistics.  But, in contrast, the northbound train appears to exhibits neither type of statistics.  In Fig.~\ref{f:619}, we display the normalized train gap histogram northbound trains at station 619 which exhibits highly-significant Poisson statistics.

\begin{figure}[t]
\includegraphics[width = .9\linewidth]{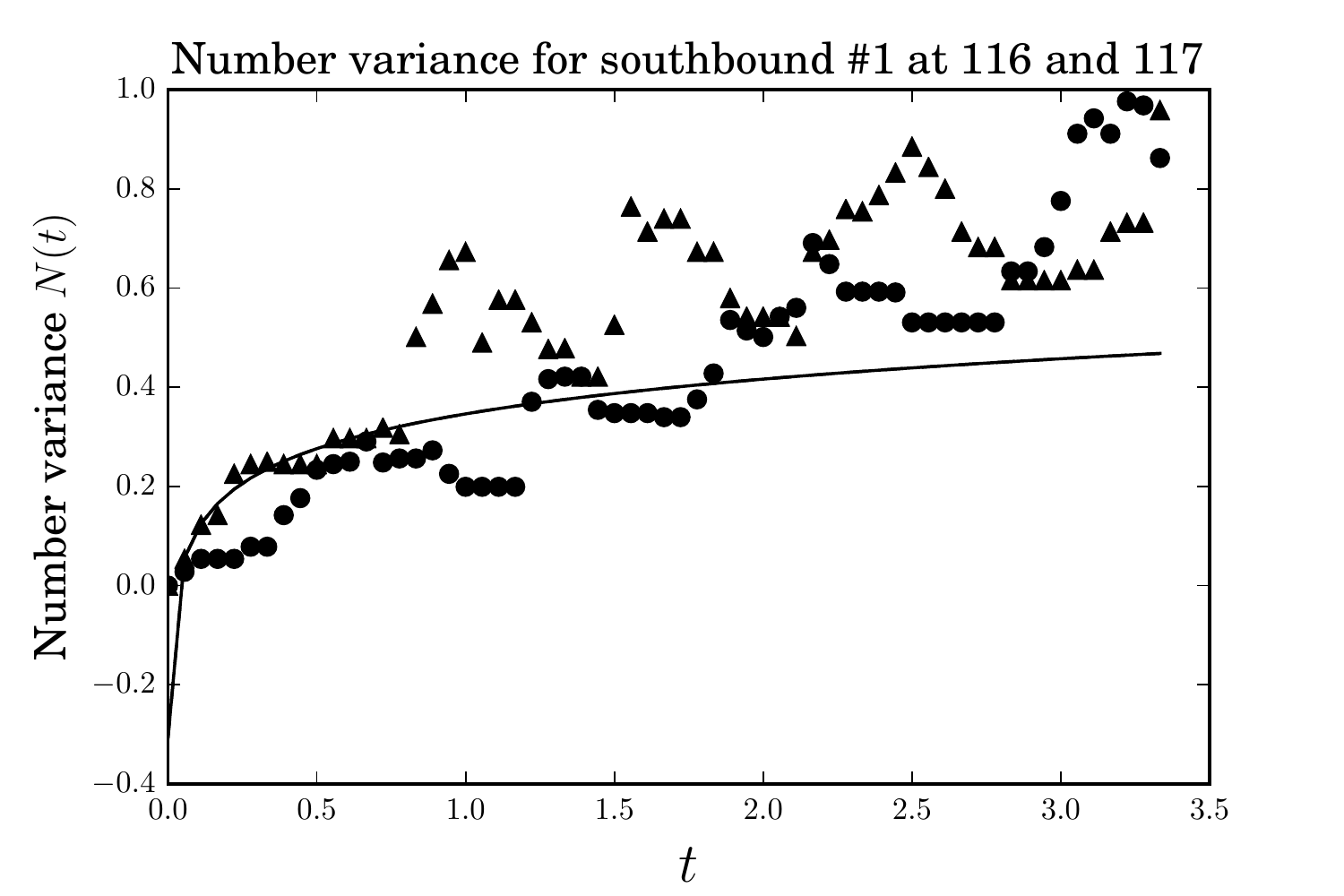}
\caption{\label{f:num_var} The empirical number variance for southbound \#1 trains at stations 116 (dots) and 117 (triangles) plotted against the theoretical curve \eqref{e:num_var}.  Agreement appears particularly good for station 117 for small values of $t$.}
\end{figure}

\paragraph{Number variance.}  To compute the number variance \eqref{e:num_var_comp}, we must obtain independent samples of the number of trains that arrive in a given time window.  We record the arrivals of southbound \#1 trains at stations 116 and 117 between 9:00am and 9:20am on weekdays.  Our data limits us to 39 samples of $n(T)$.  We plot the number variance against the theoretical prediction \eqref{e:num_var} in Fig.~\ref{f:num_var}.  While our agreement is not as good as that in \cite{Krbalek2000}, station 117 has good agreement for small values of $t$.

\begin{figure}[tbh]
\includegraphics[width = .9\linewidth]{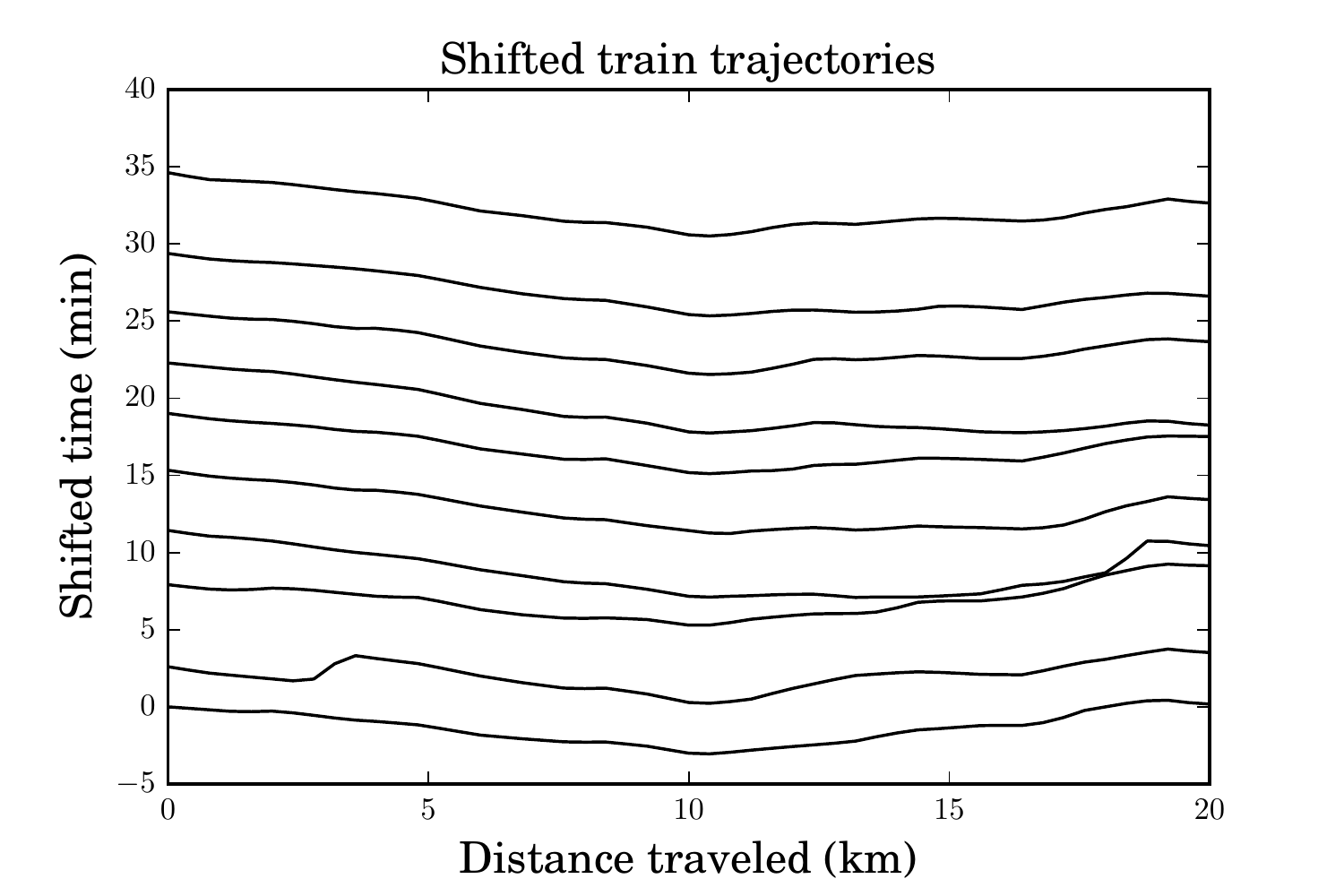}
\caption{\label{f:traj} Trajectories of the shifted southbound \#1 trains $\mu_j(\ell)$, $j = 1,2,\ldots,10$.  The horizontal axis represents the distance the train has traveled (measured from stop 103).  Theses shifted trajectories are qualitatively similar to that of non-intersecting Dyson Brownian motion, at least for short distances. }
\end{figure}

\begin{figure}[tbh]
\includegraphics[width = .9\linewidth]{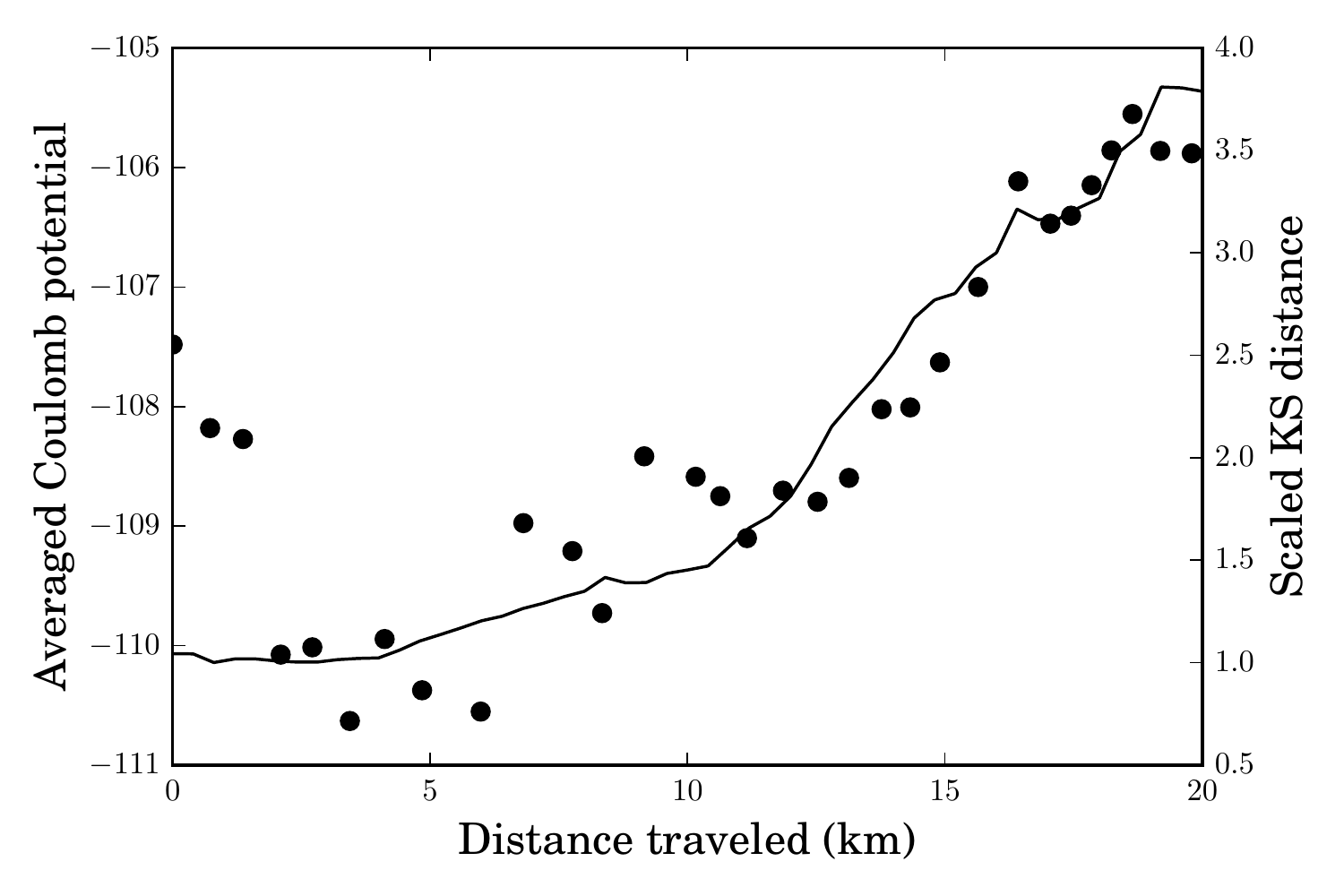}
\caption{\label{f:Coulomb} The averaged Coulomb potential \eqref{e:Coulomb} for southbound \#1 trains plotted as a function of distance from station 103. The scaled KS distance from WS is also plotted to show that when the Coulomb potential increases, so does the scaled KS statistic, indicating increased deviation from RMT statistics. }
\end{figure}

\paragraph{The Coulomb potential.}  The stationary distribution for an appropriately-scaled ($\beta = 2$) Dyson Brownian motion is the distribution on the eigenvalues $\lambda_1 < \lambda_2 < \cdots < \lambda_N$ of a GUE matrix \cite{Dyson1962}.  The Hamiltonian $H(\mathbb \lambda) : =  \frac 1 2 \sum_k \lambda_k^2 - \frac{1}{N} \sum_{j < k} \log|\lambda_k - \lambda_j|$ is approximately conserved by the Dyson Brownian motion dynamics --- the particle system $\mathbb \lambda$ fluctuates near the minimum of this functional.  The first term is referred to as the \emph{confining potential}. Given the comprehensive information our data set gives us about the MTA system we can plot many train trajectores simultaneously.  Each train is represented by a function, $\lambda_j(\ell)$, of the distance, $\ell$, the train has traveled down the track.  The value of $\lambda_j(\ell)$ is the time at which the train is a distance $\ell$ from its starting location.  This is feasible using the latitude and longitude coordinates provided by the MTA for each station.

Each weekday, we monitor 10 successive southbound \#1 trains $\lambda_j(\ell)$, $j =1,2,\ldots,10$, $0 \leq \ell \leq L$, starting with the first train ($j = 1$) that arrives at station 103 after 8am.  Each train is tracked until it reaches station 139.  For each realization of these 10 trains define
$$
\mu_j (\ell) = \lambda_j(\ell) - 90j - \langle \mathbf \lambda(L)  - \lambda(0)\rangle_j \frac{\ell}{L}
$$
where the sample average $\langle \cdot \rangle_j$ is taken over $j$.  This is used to estimate the ``velocity'' of the trains.  Define the modified Coulomb potential
\begin{align} \label{e:Coulomb}
C(\mathbf \mu(\ell) ) = \sum_{j < k} \log |\mu_k(\ell) - \mu_j(\ell)|.
\end{align}
Here we drop the confining potential. We assume we are viewing the particle system on a microscopic scale and this potential is effectively constant. In Fig.~\ref{f:traj} we plot the trajectories of $\mu_j(\ell)$ as a function of $\ell$ to demonstrate that the trains undergo non-intersecting motion.  In Fig.~\ref{f:Coulomb} we plot the averaged Coulomb potential $\langle C(\mathbf \mu(\ell) \rangle$, averaging over 29 weekdays \footnote{Ten days were rejected because at least one of the chosen trains did not complete its route.}.  The plot shows that the increase in the Coulomb potential is highly correlated with a larger scaled KS statistic. We can conjecture where the train statistics might be given by RMT based on the value of the Coulomb potential, presenting yet another connection to RMT.

It is worth pointing out in Fig.~\ref{f:Coulomb} that stations at a small distance fail the KS test but have a small Coulomb potential.  This is largely from the fact that the fluctuations of the gaps are too concentrated about their means to agree with the WS.

\paragraph{Conclusion.} In summary, we have provided significant statistical evidence that the train gaps in the NYC MTA system exhibit random matrix statistics.  In addition, regimes exists where train arrivals are Poissonian. The MTA is a concrete physical system that exhibits both RMT and Poisson statistics. We have also used detailed spatial information to gain increased insight into the train correlations, treating their trajectories as that of a particle system. While we make no conjectures about the physical mechanisms behind the transition from RMT statistics to Poissonian statistics, RMT statistics do appear to be destroyed as the train makes more and more stops.  But if one takes RMT statistics for train arrivals to be a hallmark of efficiency, as could be argued from the Cuernavaca, Mexico case study, this type of analysis may prove fruitful as a guide to understand and improve the performance of a subway system.







\bibliography{library}

\begin{table*}
\begin{tabular}{ll|ll}
\input StationKey.tex
\end{tabular}
\end{table*}

\end{document}

%% file: StationKey.tex
Station \#  & Station Name		&	Station \# &  Station Name\\
\hline
101&        Van Cortlandt Park - 242 St	&	601 &       Pelham Bay Park\\
103&        238 St			&	602  &      Buhre Av\\
104&        231 St		&		603   &     Middletown Rd\\
106&        Marble Hill - 225 St&			604&        Westchester Sq - E Tremont Av\\
107&        215 St		&		606 &       Zerega Av\\
108&        207 St		&		607  &      Castle Hill Av\\
109&        Dyckman St		&		608   &     Parkchester\\
110&        191 St		&		609    &    St Lawrence Av\\
111&        181 St		&		610     &   Morrison Av- Sound View\\
112&        168 St - Washington Hts&		611&        Elder Av\\
113&        157 St		&		612 &       Whitlock Av\\
114&        145 St		&		613  &      Hunts Point Av\\
115&        137 St - City College&		614   &     Longwood Av\\
116&        125 St		&		615    &    E 149 St\\
117&        116 St - Columbia University&		616&        E 143 St - St Mary's St\\
118&        Cathedral Pkwy		&	617&        Cypress Av\\
119&        103 St			&	618 &       Brook Av\\
120&        96 St			&	619  &      3 Av - 138 St\\
121&        86 St			&	621   &     125 St\\
122&        79 St			&	622    &    116 St\\
123&        72 St			&	623     &   110 St\\
124&        66 St - Lincoln Center	&	624      &  103 St\\
125&        59 St - Columbus Circle	&	625 &       96 St\\
126&        50 St			&	626  &      86 St\\
127&        Times Sq - 42 St		&	627   &     77 St\\
128&        34 St - Penn Station	&		628&        68 St - Hunter College\\
129&        28 St			&	629 &       59 St\\
130&        23 St			&	630  &      51 St\\
131&        18 St			&	631   &     Grand Central - 42 St\\
132&        14 St			&	632    &    33 St\\
133&        Christopher St - Sheridan Sq&		633&        28 St\\
134&        Houston St			&	634 &       23 St\\
135&        Canal St			&	635  &      14 St - Union Sq\\
136&        Franklin St			&	636   &     Astor Pl\\
137&        Chambers St			&	637    &    Bleecker St\\
138&        Cortlandt St		&		638&        Spring St\\
139&        Rector St			&	639    &    Canal St\\
140&        South Ferry Loop		&	640    &    Brooklyn Bridge - City Hall\\